\documentclass[pre,11,amsmath, twocolumn, showpacs]{revtex4-1}
\usepackage{graphicx}

\begin{document}

\title{Quantum Trajectory Approach to the Stochastic Thermodynamics of a Forced Harmonic Oscillator}

\author{Jordan M. Horowitz}

\affiliation{Departamento de F\'isica At\'omica, Molecular y Nuclear and GISC, Universidad Complutense de Madrid, 28040 Madrid, Spain}

\date{\today}

\begin{abstract}

I formulate a quantum stochastic thermodynamics for the quantum trajectories of a continuously-monitored forced harmonic oscillator coupled to a thermal reservoir.
Consistent trajectory-dependent definitions are introduced for work, heat, and entropy, through engineering the thermal reservoir from a sequence of two-level systems.
Within this formalism the connection between irreversibility and entropy production is analyzed and confirmed by proving a detailed fluctuation theorem for quantum trajectories.
Finally, possible experimental verifications are discussed.

\end{abstract}


\maketitle

\section{Introduction}

Thermal fluctuations cause the motion of a small classical system, like a colloidal particle immersed in a viscous fluid, to be random and erratic.
Such systems are clearly not within the scope of macroscopic thermodynamics \cite{Callen}.
Nevertheless, thermodynamic quantities -- such as work, heat, and entropy -- can be defined consistently along individual stochastic trajectories within the theoretical framework of stochastic thermodynamics~\cite{Seifert2005b,Schmiedl2007,Schmiedl2007b,Esposito2007b,Seifert2008,Esposito2010}.
Its two primary components are   a first-law-like energy balance equation, introduced by Sekimoto in the context of stochastic energetics \cite{Sekimoto1998, Sekimoto2007,Sekimoto}, and a definition of entropy along single fluctuating trajectories.
The predictions of stochastic thermodynamics, which have been verified experimentally~\cite{Tietz2006,Blickle2006,Speck2007}, have been important in categorizing the fluctuation theorems (reviewed in Refs.~\cite{Harris2007, Jarzynski2007,Sevick2008,Seifert2008,Jarzynski2011}) 
and in sharpening our understanding of thermodynamics at the nanoscale, especially with respect to the second law of thermodynamics~\cite{Seifert2008,Jarzynski2011}. 

Despite its significance, stochastic thermodynamics has yet to be extended to quantum mechanical systems.
Like its classical counterpart, a quantum stochastic thermodynamics would be beneficial for analyzing the fluctuations of thermodynamic quantities.
In particular, a quantum stochastic thermodynamics could provide insight into quantum extensions of the work fluctuation relations \cite{Esposito2009,Campisi2010b}: a collection of predictions regarding the fluctuations in the work performed on a  system driven away from equilibrium \cite{Bochkov1977a,Jarzynski1997a, Crooks1998,Crooks2000, Kawai2007,Jarzynski2006a,Jarzynski2007, Gomez-Marin2008a, Vaikuntanathan2009,Jarzynski2011}. 
Derivations of quantum work fluctuation relations begin with a definition of work and a method to measure its fluctuations.
For closed quantum systems -- systems that do not exchange energy with their surroundings \footnote{This terminology differs from that commonly used in thermodynamics, for example in Ref.~\cite{Esposito2009}, where a closed system exchanges energy with its environment, but not matter. The definition of closed (and open) systems used in the present article conforms to the standard usage in quantum optics~\cite{Breuer}.} -- the work performed by a quantum system during a thermodynamic process is determined by measuring the system's initial energy $U$ and final energy $U^\prime$; the difference is identified as the work $W=\Delta U=U^\prime -U$~\cite{Tasaki2000,Mukamel2003a,Talkner2007,Talkner2007b,Talkner2008b,Campisi2009,Esposito2009,Campisi2010c}.
Studies of various models have confirmed this quantum work fluctuation relation for closed quantum systems~\cite{Teifel2007,Engel2007,Talkner2008,Campisi2009b,Campisi2011} and have pointed to its limitations~\cite{Teifel2010}.
However, these predictions have never been verified experimentally, and only two experiments have been proposed \cite{Huber2008,Campisi2010b}.
When the quantum system is open -- exchanges energy with a thermal reservoir -- there are a number of proposals for the definition of work.
A common procedure is to measure the energy of the thermal reservoir  at the beginning $\epsilon$ and at the end $\epsilon^\prime$ of a thermodynamic process in addition to measuring the system's initial and final energies~\cite{Crooks2008,Talkner2009,Esposito2009,Campisi2010,Campisi2010b}.
The change in the energy of the thermal reservoir is identified as minus the heat absorbed by the system, $Q=-\Delta \epsilon=-(\epsilon^\prime-\epsilon)$, and the work is inferred from the relation $W=\Delta U-Q$.
However, the possibility of experimentally realizing a measurement of the energy of an infinite thermal reservoir is still an open question~\cite{Ritort2009,Campisi2010b}.
Esposito and Mukamel derived a quantum work fluctuation relation for an open quantum system defining the work along trajectories constructed from formal solutions of a quantum master equation \cite{Esposito2006}, though the connection between the work defined in Ref.~\cite{Esposito2006} and a measurable physical quantity remains unclear.
In the high-friction limit, Deffner and Lutz investigated the work performed by an open quantum system~\cite{Deffner2008,Deffner2009,Deffner2010}.
However, in this limit the equations of motion are effectively classical, as there are no quantum coherences (yet quantum fluctuations remain in addition to the thermal fluctuations).
A quantum work fluctuation theorem has also been predicted by De Roeck and Maes for a path-dependent work defined along unitary quantum evolutions interrupted by projective measurements~\cite{DeRoeck2004}.
Yet, another approach was put forward recently by Hu and Subi\c{s}i using a decoherent histories analysis of a non-markovian quantum brownian motion~\cite{Hu2012}.
Despite the many studies of work relations in open quantum systems, there are still unresolved questions regarding the possibility of experimentally measuring the work in a coherent quantum system, such questions that  could be resolved by constructing a quantum stochastic thermodynamics.

In this article, I take a first step in formulating a quantum stochastic thermodynamics using the quantum trajectory formalism, originally developed in the field of quantum optics in response to experiments on continuously-monitored individual quantum systems~\cite{Carmichael,Wiseman1994,Plenio1998,Jacobs1998,Jacobs2006,Brun2002,Breuer}.
Within this formalism quantum trajectories are the stochastic evolution of a quantum system conditioned on the outcomes of a sequence of weak measurements~\cite{Wiseman1994}.
For concreteness, I analyze a specific model inspired by current experiments in cavity quantum electrodynamics~\cite{Raimond2001,Walther2006,Gleyzes2007}:  a forced quantum harmonic oscillator weakly coupled to a thermal reservoir.
However, instead of considering an infinite thermal reservoir, the thermal reservoir is engineered by coupling the harmonic oscillator with a rapid succession of two-level systems, one at a time~\cite{Kist1999,Santos2011}.
By measuring the state of each two-level system after its interaction with the harmonic oscillator, we are able to continuously monitor the evolution of the harmonic oscillator and measure the amount of energy transferred to the thermal reservoir, a strategy originally proposed by Crooks \cite{Crooks2008b,Crooks2008}.
Derezi\'nski, De Roeck, and Maes have also proposed the use of quantum trajectories in the study of quantum fluctuation theorems for currents, but did not comment on their experimental realizability~\cite{DeRoeck2004,DeRoeck2006}.
By studying a concrete model, I am able to devise experimentally verifiable definitions of thermodynamic quantities along individual quantum trajectories.

We will see that  because the forced quantum harmonic oscillator's evolution is perturbed by its interaction with the sequence of two-level systems, its effective equation of motion is a stochastic Schr\"odinger equation.
The derivation presented in this article will follow the approach developed in Refs.~\cite{Dalibard1992,Kist1999,Santos2011,Ueda1992}, though adapted for a time-dependent Hamiltonian.
This approach to open quantum systems, utilizing repeated interactions, has been studied rigorously by Attal and Pautrat~\cite{Attal2006} and Attal and Joye~\cite{Attal2007} employing the quantum stochastic calculus devised by Parsatharathy~\cite{Parsatharathy}.
While the present analysis does not utilize Parsatharathy's quantum stochastic calculus, Pellegrini~\cite{Pellegrini2010} and later Attal and Pellegrini~\cite{Attal2010} demonstrated that the stochastic Schr\"{o}dinger equation considered here can be naturally incorporated into this framework.
Stochastic Schr\"odinger equations have also appeared in dynamic reduction models, where they are postulated as a means to dynamically localize a state vector in Hilbert space~\cite{Bassi2003}.
The novelty of the present endeavor is to consider explicitly time-dependent Hamiltonians.
By applying the secular approximation to the full time-dependent Hamiltonian (without utilizing Floquet theory as in Ref.~\cite{Breuer}), I derive a stochastic Schr\"{o}dinger equation and a quantum master equation valid even for nonadiabatic driving; a result that is necessary as we are interested far-from-equilibrium quantum thermodynamics.

The construction of a quantum stochastic thermodynamics begins in Sec.~\ref{sec:model} by specifying the model and introducing the stochastic Schr\"odinger equation.
Then in Sec.~\ref{sec:stochThermo}, I formulate a quantum stochastic thermodynamics by defining heat, work, and entropy along individual quantum trajectories.
Work fluctuation relations are analyzed in Sec.~\ref{sec:flucRe}, where a detailed fluctuation relation for quantum trajectories is derived.
Finally, I conclude in Sec.~\ref{sec:discussion} with a discussion of possible experimental verifications of quantum stochastic thermodynamics in cavity quantum electrodynamics experiments~\cite{Raimond2001,Walther2006,Gleyzes2007}.

\section{Forced Quantum Harmonic Oscillator Coupled to a Thermal Reservoir}\label{sec:model}

In this section, I specify the details of the model and fix notation.
The model is a one-dimensional forced harmonic oscillator weakly coupled to a thermal reservoir.
Since a common experimental realization of the quantum harmonic oscillator is a single mode of an electromagentic field in a superconducting microwave cavity~\cite{Raimond2001,Walther2006,Gleyzes2007}, the terminology and techniques of quantum optics will prove helpful in developing and explaining the model.
The thermal reservoir is engineered by weakly coupling the harmonic oscillator with a succession of two-level systems.
By continually measuring their outgoing states, we are able to continuously monitor the evolution of the oscillator.
The effective equation of motion governing the dynamics of the continuously-monitored oscillator is a stochastic Schr\"odinger equation (Eq.~\ref{eq:sse} below), which is presented in this section.
Its derivation, though a basic extension of the techniques utilized in Refs.~\cite{Kist1999,Santos2011,Ueda1992}, is technical and  is therefore reserved for Appendix~\ref{sec:derivation}.

\begin{figure}[ht]
\centering
\includegraphics[scale=0.25,angle=-90]{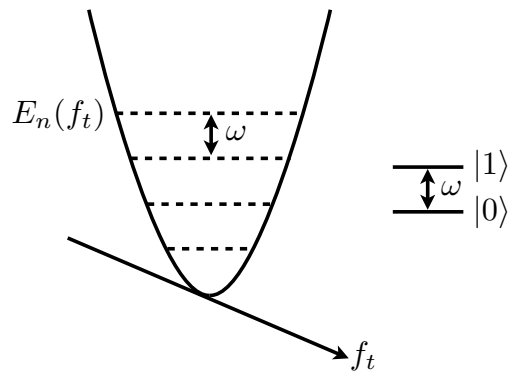}
\caption{Depiction of the energy levels $E_n(f_t)$ for the instantaneous eigenstates $|n_{f_t}\rangle$ of a forced quantum harmonic oscillator of frequency $\omega$ with forcing protocol $f_t$.
The oscillator interacts resonantly with a sequence of two-level systems with ground states $|0\rangle$ and excited states $|1\rangle$ in order to simulate a thermal reservoir at inverse temperature $\beta$.
}
\label{fig:osc}
\end{figure}

I have in mind a forced quantum harmonic oscillator of mass $m$ and frequency $\omega$, with position $x$ and momentum $p$, coupled to a thermal reservoir at inverse temperature $\beta$, as depicted in Fig.~\ref{fig:osc}.
The Hamiltonian of the harmonic oscillator,
\begin{equation}\label{eq:H2}
\begin{split}
H(f)&=\frac{p^2}{2m}+\frac{1}{2}m\omega^2x^2-m\omega^2 f x \\
&=\frac{p^2}{2m}+\frac{1}{2}m\omega^2(x-f)^2-\frac{1}{2}m\omega^2f^2,
\end{split}
\end{equation}
is parameterized by an externally controlled force, or parameter, $f$, which is varied with time in order to do work on the oscillator.
Here, and throughout the following $\hbar=1$.
For an electromagnetic field, the forcing can be induced by deterministically varying with time a collection of macroscopic classical sources~\cite{Cohen}.
As can be inferred from Eq.~\ref{eq:H2}, a harmonic oscillator in the presence of an external force is equivalent to a harmonic oscillator that has been translated horizontally a distance $f$, and whose energy has been lowered by $(1/2)m\omega^2f^2$.
This observation motivates the definition of lowering and raising operators parameterized by~$f$,
\begin{equation} \label{eq:a}
\begin{split}
a_f &= \sqrt{\frac{m\omega}{2}}\left(x-f+\frac{ip}{m\omega}\right)\\ 
a_f^\dag &= \sqrt{\frac{m\omega}{2}}\left(x-f-\frac{ip}{m\omega}\right),
\end{split}
\end{equation}
which when substituted into the Hamiltonian [Eq.~\ref{eq:H2}] lead to the simplified expression
\begin{equation}
\label{eq:oscHam}
H(f)=\omega\left(a_f^\dag a_f+\frac{1}{2}\right)-\frac{1}{2}m\omega^2f^2.
\end{equation}
For each $f$, the eigenstates  $|n_f\rangle$ of the number operator $N_f=a_f^\dag a_f$, with eigenvalues $n$, are also eigenstates of the Hamiltonian $H(f)$ with energies
\begin{equation}
E_n(f)=\omega\left(n+\frac{1}{2}\right)-\frac{1}{2}m\omega^2f^2.
\end{equation}

From time $t=0$ to $\tau$, we drive the oscillator by varying the force $f$ using the linear protocol ${\mathcal F}=\{f_t\}_{t=0}^\tau$, where $f_t=\sqrt{2/(m\omega)}\nu t$ and $\nu$ is the rate at which the force varies.
For notational simplicity, I will denote the lowering and raising operators [Eq.~\ref{eq:a}] at time $t$ when the force has the value $f_t$ as $a_t= a_{f_t}$ and $a_t^\dag= a_{f_t}^\dag$.
These operators, $a_t$ and $a_t^\dag$, correspond to the lowering and raising operators for the eigenbasis $|n_{f_t}\rangle$ that at each $t$ diagonalizes the Hamiltonian: $H(f_t)|n_{f_t}\rangle=E_n(f_t)|n_{f_t}\rangle$.
These states $|n_{f_t}\rangle$ -- which are not solutions of the Schr\"{o}dinger equation -- I will call the instantaneous eigenstates of $H(f_t)$, though sometimes they are refereed to as the adiabatic basis~\cite{Mukamel2003a} due their employment in the derivation of the quantum adiabatic theorem.

\begin{figure}[ht]
\centering
\includegraphics[scale=0.3, angle=-90]{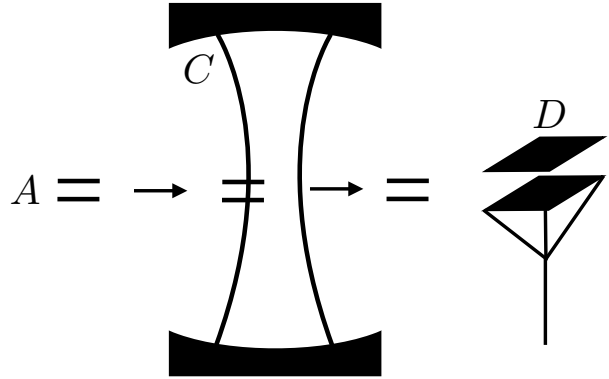}
\caption{Illustration of a possible experimental setup realizing a forced harmonic oscillator weakly coupled to a thermal reservoir.
A sequence of two level-atoms, $A$, pass through a cavity $C$, after which their final state is measured by detector $D$.}
\label{fig:osc2}
\end{figure}

As we drive the oscillator, it will continually exchange energy with a thermal reservoir at inverse temperature $\beta$.
However, I will not model the thermal reservoir as very large number of degrees of freedom.
Instead, I follow Refs.~\cite{Pielawa2010,Kist1999,Santos2011}  and engineer the thermal reservoir by weakly coupling the harmonic oscillator to a sequence of two-level systems, which I will call atoms since this procedure was originally proposed in the context of cavity quantum electrodynamics~\cite{Raimond2001,Walther2006,Gleyzes2007}.
A possible experimental realization in cavity quantum electrodynamics is illustrated in Fig.~\ref{fig:osc2}.
The Hamiltonian of each atom is
\begin{equation}\label{eq:atomHam}
H_A=\omega \sigma^\dag\sigma,
\end{equation}
where $\sigma=|0\rangle\langle1|$ is the atomic lowering operator inducing transitions from the atom's excited state $|1\rangle$ to ground state $|0\rangle$.
Prior to interacting with the oscillator, each atom is prepared in a definite known state: either its ground state $|0\rangle$ with probability 
\begin{equation}\label{eq:r0}
r_0=\frac{1}{1+e^{-\beta\omega}}
\end{equation}
or its excited state $|1\rangle$ with probability 
\begin{equation}\label{eq:r1}
r_1=\frac{e^{-\beta\omega}}{1+e^{-\beta\omega}}.
\end{equation}
Then, one at a time, each atom interacts with the oscillator for a short time $\delta t$, after which we make an instantaneous projective measurement of the atom's state in order to determine whether it remained in its initial state or transitioned.
By measuring the states of the atoms after they interact with the oscillator, we are able to track (or monitor) the  oscillator's evolution.
Moreover, knowledge of whether the atom transitioned or not, allows us to monitor how much energy is transfered to the atom during its interaction with the oscillator, and consequently how much energy has been exchanged with the thermal reservoir. 

The coupling between each atom and the oscillator at time $t$ is mediated by the displaced raising and lowering operator,
\begin{equation}\label{eq:bara}
\begin{split}
{\bar a}^\dag_f&=a_f^\dag+i\nu/\omega\\
{\bar a}_f&=a_f-i\nu/\omega,
\end{split}
\end{equation}
through the interaction Hamiltonian
\begin{equation}\label{eq:HI}
H_I(f_t)=\lambda({\bar a}_t^\dag\sigma+{\bar a}_t\sigma^\dag),
\end{equation}
with weak-coupling, $\lambda\ll\omega$.
In the absence of driving ($\nu=0$), this interaction Hamiltonian is the Jaynes-Cummings Hamiltonian, known to quantitatively describe experimental observations in microwave cavities~\cite{Breuer,Raimond2001}.
However, in the language of quantum optics, the lowering and raising operators of the harmonic oscillator are displaced by $i\nu/\omega$ due to the presence of the classical field induced by the motion of the classical sources.
In addition, $a_t$ and $a^\dag_t$ are evaluated at $t$, inducing transitions between the instantaneous eigenstates of $H(f_t)$, which is vital for interpreting the sequence of atoms as a thermal reservoir constantly exchanging energy with the oscillator.
A rigorous justification for the form of $H_I$ in Eq.~\ref{eq:HI} is presented in Appendix~\ref{sec:rotwave}.
There a two-time scale perturbation analysis is applied to the Schr\"odinger equation for the oscillator coupled to an atom in the dipole approximation.
The analysis indicates that the dynamics are well approximated by $H_I(f_t)$~[Eq.~\ref{eq:HI}] on times $\lambda \delta t = O(1)$ when the coupling is weak, $\lambda \ll \omega$.

Having introduced the model, I now present the stochastic Schr\"odinger equation governing the evolution of the harmonic oscillator conditioned on the measurement outcomes~\cite{Wiseman1994,Jacobs1998,Jacobs2006,Brun2002,Breuer}.
The stochastic Schr\"odinger equation (Eq.~\ref{eq:sse} below) is an effective equation of motion for the evolution of the harmonic oscillator state vector $|\psi_t\rangle$ at time $t$ conditioned on all the measurements made on the atoms prior to time $t$. 
In its derivation, I have assumed that (i) each atom interacts for the same amount of time $\delta t$; (ii) during each $\delta t$ only one atom interacts with the oscillator; (iii) the interaction time is short, $\lambda \delta t\ll1$; (iv) the interaction is weak $\lambda \ll \omega$; (v) the mean number of excitations in the oscillator ${\bar n}_t$ at each time $t$ is small, $\lambda\delta t {\bar n}_t\ll1$; and (vi) the driving is not too fast, $\lambda \delta t (\nu/\omega)\ll1$.
While this last requirement restricts the rate of the driving $\nu$, it does not necessitate the adiabatic approximation ($\nu\ll\omega$): assumption (vi) can still be satisfied with $\nu\sim\omega$.

Consider the evolution of the harmonic oscillator during a small time interval $dt$ large compared to $\delta t$.
During $dt$ many atoms interact with the oscillator.
Prior to interacting with the oscillator, each atom is prepared in a known state, either $|0\rangle$ or $|1\rangle$.
As I demonstrate in Appendix~\ref{sec:derivation}, the probability during $dt$ to measure an outgoing state  different from an initial state -- that is to observe a transition or jump -- is small. 
In particular, the probability to observe a jump from $|0\rangle$ to $|1\rangle$ during $dt$ is $p_{01}=gr_0\langle {\bar a}^\dag_t {\bar a}_t\rangle_tdt$, where $g=\lambda^2\delta t$ is the jump rate and $\langle \cdot \rangle_t=\langle \psi_t|\cdot|\psi_t\rangle$ denotes the quantum mechanical expectation with respect to the conditioned oscillator state vector at $t$;
the probability to jump from $|1\rangle$ to $|0\rangle$ during $dt$ is $p_{10}=gr_1\langle {\bar a}_t {\bar a}^\dag_t\rangle_tdt$;
and the probability of observing no transitions is $1-p_{01}-p_{10}$.
Since the probability to observe a jump is of order $dt$, the stochastic sequence of jumps is a Poisson process.
Mathematically, I introduce two stochastic Poisson increments corresponding to an atom jumping up (the system jumping down),  $dN_t^+$, and an atom jumping down (the system jumping up), $dN_t^-$.
These Poisson increments are a sequence of random numbers that are either $0$ or $1$ -- 
\begin{equation}
\left(dN_t^+\right)^2=dN^+_t \quad\textrm{and}\quad \left(dN_t^-\right)^2=dN^-_t
\end{equation}
 -- whose ensemble expectation value at $t$ is 
\begin{equation}\label{eq:dN}
\begin{split}
E[dN_t^+]&=gr_0\langle {\bar a}^\dag_t{\bar a}_t\rangle_tdt, \\
E[dN_t^-]&=gr_1\langle {\bar a}_t{\bar a}^\dag_t\rangle_tdt.
\end{split}
\end{equation}
The Poisson increments allow us to compactly write the stochastic Schr\"odinger equation for the stochastic differential of the conditional state vector during the time interval $[t,t+dt)$, $d|\psi_t\rangle=|\psi_{t+dt}
\rangle-|\psi_t\rangle$, as the It\={o} stochastic differential equation
\begin{align}
\label{eq:sse}
\nonumber
d|\psi_t\rangle=&dt\Big(-iH(f_t)-\frac{gr_0}{2}{\bar a}^\dag_t{\bar a}_t-\frac{gr_1}{2}{\bar a}_t{\bar a}^\dag_t  \\ \nonumber
& +\frac{gr_0}{2}\langle {\bar a}^\dag_t{\bar a}_t\rangle_t+\frac{gr_1}{2}\langle {\bar a}_t{\bar a}^\dag_t\rangle_t\Big)|\psi_t\rangle \\ \nonumber
&+dN^-_t\left(\frac{{\bar a}^\dag_t|\psi_t\rangle}{\sqrt{\langle {\bar a}_t {\bar a}_t^\dag \rangle_t}} -|\psi_t\rangle\right) \\
&+dN^+_t\left(\frac{{\bar a}_t|\psi_t\rangle}{\sqrt{\langle {\bar a}^\dag_t {\bar a}_t \rangle_t}} -|\psi_t\rangle\right).
\end{align}
Most of the time the stochastic increments $dN^j_t$, $j=+,-$, in Eq.~\ref{eq:sse} are zero and $|\psi_t\rangle$ evolves deterministically under the action of the term proportional to $dt$ in Eq.~\ref{eq:sse}.
Occasionally a jump is detected, signaling an abrupt change in the oscillator's state vector.
When an atom jumps up $dN^+_t$ equals one, and the oscillator's state vector experiences a discontinuous change mediated by ${\bar a}_t$.
Similarly, when $dN^-_t$ is one, the oscillator changes abruptly under the action of ${\bar a}^\dag_t$. 


The state vector $|\psi_t\rangle$ characterizes the sub-ensemble of systems conditioned on a particular sequence of measurement outcomes.
Because each measurement outcome is random, the conditioned sub-ensemble and its corresponding state vector $|\psi_t\rangle$ vary stochastically.
Equation~\ref{eq:sse} is the equation of motion describing this stochastic evolution of $|\psi_t\rangle$ through Hilbert space \cite{Breuer}.
The goal in formulating a quantum stochastic thermodynamics will be to associate with the stochastic evolution of the state vector and its corresponding ensemble notions of work, heat, and entropy.

The state vector $|\psi_t\rangle$ describes the conditional (or selective) evolution of the oscillator.
When the measurement outcomes are ignored (or averaged over), the unconditional evolution is characterized by the density matrix obtained as the average over all measurements~\cite{Breuer},
\begin{equation}\label{eq:rho}
\rho_t=E[|\psi_t\rangle\langle\psi_t|].
\end{equation}
The equation of motion for the density matrix  $\rho_t$, calculated by differentiating Eq.~\ref{eq:rho} with the aid of Eq.~\ref{eq:sse}, is the Linblad master equation~\cite{Breuer},
\begin{equation}\label{eq:master}
\partial_t\rho_t=-i[H(f_t),\rho_t]+gr_0{\mathcal D}[{\bar a}_t]\rho_t + gr_1 {\mathcal D}[{\bar a}_t^\dag]\rho_t,
\end{equation}
where
\begin{equation}
{\mathcal D}[c]\rho=c\rho c^\dag-\frac{1}{2}c^\dag c\rho-\frac{1}{2}\rho c^\dag c.
\end{equation}
For fixed $f$ ($\nu=0$), Eq.~\ref{eq:master} is a master equation describing the evolution of a harmonic oscillator in the presence of a thermal reservoir at the inverse temperature $\beta$, whose value can be inferred from the ratio of Eqs.~\ref{eq:r0} and \ref{eq:r1} \cite{Breuer},
\begin{equation}\label{eq:Boltzmann}
\frac{r_1}{r_0}=e^{-\beta\omega}.
\end{equation}
In particular, if we allow the system to evolve freely with $f$ fixed, in the long-time limit the system will relax to equilibrium with a density matrix given by the Boltzmann density matrix
\begin{equation}\label{eq:eqDens}
\rho^{eq}_f=\frac{e^{-\beta H(f)}}{Z_f},
\end{equation}
where $Z_f={\rm Tr}[\exp(-\beta H(f))]$ is the partition function.
Thus, the average evolution of the harmonic oscillator induced by its interaction with a sequence of two-level atoms is equivalent to the evolution caused by a thermal reservoir.
In general, when $\nu\neq0$, Eq.~\ref{eq:master} describes the nonequilibrium evolution of the oscillator.
Specifically, the solution to the master equation~[Eq.~\ref{eq:master}] at $t$, $\rho_t$, may differ from the instantaneous equilibrium density matrix $\rho^{eq}_{f_t}$.

\section{Quantum Stochastic Thermodynamics}\label{sec:stochThermo}

In this section, I introduce a quantum stochastic thermodynamics for the forced harmonic oscilator outlined in Sec.~\ref{sec:model}.
Stochastic thermodynamics, like its classical counterpart, will be formulated along thermodynamic processes, which are introduced in Sec.~\ref{sec:thermoProc}.
Then in Sec.~\ref{sec:firstLaw}, the change in energy along a thermodynamic process is divided into two contributions: the work and the heat.
Entropy and entropy production are identified in Sec.~\ref{sec:entropy}.

\subsection{Thermodynamic Processes}\label{sec:thermoProc}

In classical macroscopic thermodynamics, work and heat characterize the exchange of energy between a system and its surroundings during a thermodynamic process.
Such a process is specified by a sequence of macroscopic actions (or events) executed to manipulate the evolution of the system.
Typical actions taken during a thermodynamic process include preparing the system initially in thermodynamic equilibrium at fixed temperature; or modifying the volume or the pressure using a specified protocol.
Similarly, in order to construct a quantum stochastic thermodynamics, I will first specify the sequence of actions that define a thermodynamic process.
Two processes, introduced below, will prove useful: the forward and reverse processes, which are related by time reversal.

A process will be composed of four parts: an initial preparation of the system; an initial observable that is measured at the beginning of the process; a protocol for varying the external force with time; and a final observable measured upon completion of the process.
The measurements of observables at the beginning and end of the process are macroscopic actions unique to quantum processes.
They are necessary in order to guarantee that the forward and reverse processes begin in a pure state, and are required in the proof of the detailed fluctuation theorem in Sec.~\ref{sec:flucRe} below.

We begin the forward process by preparing the initial ensemble of quantum systems.
We form this ensemble by collecting a large number of quantum systems each in a known eigenstate $|a\rangle$ of an observable $A$ with non-degenerate eigenvalue $a$, such that the fraction of systems in eigenstate $|a\rangle$ is proportional to the probability $P_a$.
The statistical properties of this ensemble are characterized by the density matrix
\begin{equation}\label{eq:rhoA}
\rho_0=\rho_A=\sum_a P_a|a\rangle\langle a|.
\end{equation}
We then randomly select a quantum system from the initial ensemble $\rho_A$, which will be in state $|a\rangle$ with probability $P_a$.
To confirm this state, we make a projective measurement of $A$.
Then from time $t=0$ to $\tau$, the force is varied according to the protocol ${\mathcal F}$, while continuously monitoring the oscillator through the sequence of interacting atoms.
During this time the oscillator state vector evolves stochastically according to Eq.~\ref{eq:sse}.
In any realization, we will observe that the atoms jump at a specific sequence of times.
We record each jump time $t_k$ and the type of jump $m_k=+,-$, where $+$ denotes an atom jumping up and $-$ denotes an atom jumping down.
At time $\tau$, we fix the external force at $f_\tau=\sqrt{2/(m\omega)}\nu\tau$, cease the flow of atoms, and make a projective measurement of a new observable $B$.
The outcome is one of its non-degenerate eigenvalues $b$, corresponding to the eigenstate $|b\rangle$, with probability $P_b=|\langle b|\psi_{\tau^-}\rangle|^2$.
Consequently, at time $t=\tau$ the state vector for any realization will collapse into one of the eigenstates of $B$, and the density matrix [Eq.~\ref{eq:rho}] will be diagonal in the eigenbasis of $B$:
\begin{equation}\label{eq:rhoB}
\rho_{\tau}=\rho_B=\sum_b P_b |b\rangle\langle b|.
\end{equation}
Repeating this series of actions generates an ensemble of realizations.

During each realization of the process, we make a measurement of $A$ with outcome $a$ at the beginning of the process, a measurement of $B$ with outcome $b$ at the end of the process, and observe a series of jumps $m_k$ at times $t_k$.
I collect this string of measurement outcomes into a vector, which I call the measurement trajectory,
$\gamma=\{a; m_1,t_1; m_2, t_2; \cdots; b\}$.
Furthermore, at each time $t$ the system is characterized by the state vector conditioned on all measurements up to $t$, $|\psi_t(\gamma)\rangle$, which is the solution of Eq.~\ref{eq:sse}.
The quantum trajectory is the sequence of state vectors traced out by the system through Hilbert system during the time interval from $t=0$ to $\tau$,  denoted by $\psi(\gamma)=\{|\psi_t(\gamma)\rangle\}_{t=0}^\tau$ or simply $\psi$ when $\gamma$ is clear from the context.

The reverse process is defined as the time-reversed forward process in which each action of the forward process is carried out in the reverse order.
First, recall that time-reversal in quantum mechanics is implemented by the time-reversal operator $\Theta$, which is an antilinear -- 
\begin{equation}
\Theta c=c^*\Theta,
\end{equation}
for any complex number $c$, where $*$ denotes complex conjugation -- 
 involution, 
\begin{equation}
\Theta^2=I,
\end{equation}
where $I$ is the identity operator~\cite{Sakurai}.

In the reverse process, we prepare the initial ensemble by collecting a number of quantum systems, each in an eigenstate of the time-reveresed observable of $B$, ${\tilde B}=\Theta B\Theta^{-1}$.
The ensemble is constructed so that each eigenstate $|{\tilde b}\rangle=\Theta|b\rangle$ of ${\tilde B}$ with eigenvalue $b$ occurs with probability ${\tilde P_b}$ -- which in general differs from the probability to measure $b$ at the end of the forward process, $P_b$.
Hence, the initial ensemble is characterized by the density matrix
\begin{equation}\label{eq:rhoTilde0}
\tilde\rho_0=\tilde\rho_B=\sum_b{\tilde P}_b|{\tilde b}\rangle\langle {\tilde b}|.
\end{equation} 
We then randomly select a quantum system from this ensemble and make a projective measurement of ${\tilde B}$.
After, we drive the system from time $t=0$ to $\tau$ using the time-reversed protocol $\tilde{\mathcal F}=\{{\tilde f}_t\}_{t=0}^\tau$, where ${\tilde f}_t=f_{\tau-t}=\sqrt{2/(m\omega)}\nu(\tau-t)$.
During this time interval, the oscillator interacts with a sequence of atoms; we record the times at which the atoms jump $t_k$ and the types of jumps $m_k$, as in the forward process.
Finally, we measure the time-reversed operator ${\tilde A}=\Theta A\Theta^{-1}$ obtaining eigenvalue $a$ corresponding to eigenvector $|{\tilde a}\rangle=\Theta|a\rangle$ with probability ${\tilde P}_a=|\langle {\tilde a}|\psi_{\tau^-}\rangle|^2$.
At the time of measurement the state vector collapse into an eigenvector of ${\tilde A}$ and the density matrix becomes diagonal in the eigenbasis of ${\tilde A}$:
\begin{equation}
\tilde\rho_\tau=\tilde\rho_A=\sum_a \tilde{P}_a|{\tilde a}\rangle\langle {\tilde a}|.
\end{equation}

Every measurement trajectory of the forward process during which $M$ jumps are observed, $\gamma=\{a; m_1,t_1; m_2, t_2; \cdots; m_M,t_M; b\}$, is paired with a conjugate reverse measurement trajectory of the reverse process $\tilde\gamma=\{{\tilde b}; {\tilde m}_1,\tau-t_M; \cdots; {\tilde m}_{M-1},\tau-t_{2};{\tilde m}_M,\tau-t_1;{\tilde a}\}$, where ${\tilde m}_{M-k}$ is $+$ ($-$) when $m_{k+1}$ is $-$ ($+$), for $k=0,\dots M-1$: when an atom jumps up along a forward trajectory, the atom will jump down in the conjugate reverse trajectory.
The quantum trajectory traced out through Hilbert space by the oscillator's state vector corresponding to the measurement trajectory $\tilde{\gamma}$ will be denoted ${\tilde \psi}(\tilde \gamma)=\{|\psi_t(\tilde \gamma)\rangle\}_{t=0}^\tau$, where $|\psi_t(\tilde \gamma)\rangle$ is the solution of Eq.~\ref{eq:sse} during the reverse process.
Observe that in general the sequence of state vectors traced out by the system is not the time-reversal of the quantum trajectory observed in the forward process, $|\psi_t(\gamma)\rangle\neq\Theta|\psi_{\tau-t}(\tilde\gamma)\rangle$.

\subsection{First Law of Stochastic Thermodynamics}\label{sec:firstLaw}

In this section, I formulate a energy balance equation (Eq.~\ref{eq:work} below) relating the heat and work to the change in energy along individual quantum trajectories.  
I proceed by first identifying the heat as the energy transferred to the thermal reservoir.
The energy is then defined as the quantum mechanical ensemble average of the Hamiltonian.
Finally, the change in the energy not accounted for by heat is identified as the work.

Heat is the energy exchanged with a thermal reservoir.
Since the thermal reservoir in the present model is composed of a sequence of two-level atoms, I identify the heat absorbed by the oscillator as the energy released from the atoms.
Energy is only exchanged when an atom transitions or jumps between its energy eigenstates.
In particular, each time we observe a jump from $|0\rangle$ to $|1\rangle$, the atom has absorbed $\omega$ energy from the oscillator; similarly, when the atom has jumped from $|1\rangle$ to $|0\rangle$, it has released $\omega$ energy.
Therefore, for a given quantum trajectory $\psi(\gamma)$, the increment in the heat absorbed by the oscillator during the small time interval from $t$ to $t+dt$ may be written as
\begin{equation}\label{eq:dQ}
dQ_t[\psi]=\omega(dN^-_t-dN^+_t).
\end{equation}
The heat absorbed by the oscillator during $[0,t)$ is the stochastic integral of Eq.~\ref{eq:dQ}, 
\begin{align}\label{eq:Q}
Q_t[\psi]&=\int_0^t dQ_s[\psi] \\ 
&=\int_0^t \omega (dN^-_s-dN^+_s).
\end{align}

Next, I define the (internal) energy at time $t$ for the entire ensemble characterized by the state vector $|\psi_t(\gamma)\rangle$ conditioned on all the measurements up to $t$.
The energy at $t$ along the quantum trajectory $\psi(\gamma)$ is the ensemble average of the Hamiltonian,
\begin{equation}\label{eq:E}
U_t[\psi]=\langle \psi_t|H(f_t)|\psi_t\rangle.
\end{equation}
The energy varies with time due to the time dependence of the Hamiltonian as well as the stochastic evolution of the state vector.
In particular, the change in the energy up to $t$ is
\begin{align}\label{eq:energy}
\Delta U_t[\psi] &= \int_0^t dU_s[\psi]
\end{align}
where, $dU_t[\psi]=U_{t+dt}[\psi]-U_t[\psi]$,  is the stochastic differential of $U_t$ during the small interval from $t$ to $t+dt$:
\begin{align} \label{eq:dU}\nonumber
dU_t[\psi]= &dt {\dot f}_t\langle \psi_t|\partial_f H(f_t)|\psi_t\rangle \\ \nonumber
&-dt\, g r_0 \langle\psi_t|\left[\frac{1}{2}\{H(f_t),{\bar a}^\dag_t {\bar a}_t\}-H(f_t)\langle {\bar a}^\dag_t {\bar a}_t\rangle_t\right]|\psi_t\rangle \\ \nonumber
&-dt\, g r_1 \langle\psi_t|\left[\frac{1}{2}\{H(f_t),{\bar a}_t {\bar a}^\dag_t\}-H(f_t)\langle {\bar a}_t {\bar a}^\dag_t\rangle_t\right]|\psi_t\rangle \\ \nonumber
&+dN_t^-\left(\frac{\langle {\bar a}_tH(f_t){\bar a}^\dag_t\rangle_t}{\langle {\bar a}_t {\bar a}^\dag_t\rangle_t}-\langle H(f_t)\rangle_t\right) \\
&+dN_t^+\left(\frac{\langle {\bar a}^\dag_tH(f_t){\bar a}_t\rangle_t}{\langle {\bar a}^\dag_t {\bar a}_t\rangle_t}-\langle H(f_t)\rangle_t\right),
\end{align}
which can be deduced from the definition of $U_t$ [Eq.~\ref{eq:energy}] with the aid of Eq.~\ref{eq:sse}.
Here, for two operators $O$ and $O^\prime$, $\{O,O^\prime\}=OO^\prime+O^\prime O$ is the anti-commutator.
For a process where the initial observable is the initial Hamiltonian, $A=H(f_0)$, and at the end of the process the final Hamiltonian is measured $B=H(f_\tau)$, this definition of the change in internal energy in Eq.~\ref{eq:dU} agrees with that commonly encountered in the derivation of quantum work relations~\cite{Campisi2010b}.
However, its statistics are diferrent due to the interaction with the reservoir.

Finally, the work during the time interval from $t$ to $t+dt$ is identified as the change in the energy not accounted for by heat \cite{Sekimoto,Seifert2008}:
\begin{equation}
dW_t[\psi]=dU_t[\psi]-dQ_t[\psi],
\end{equation}
which may be formulated, using the definitions of $dU_t$~[Eq.~\ref{eq:dU}] and $dQ_t$ [Eq.~\ref{eq:dQ}], as 
\begin{align} \label{eq:dW}\nonumber
dW_t[\psi]= &dt {\dot f}_t\langle \psi_t|\partial_f H(f_t)|\psi_t\rangle \\ \nonumber
&-dt\, g r_0 \langle\psi_t|\left[\frac{1}{2}\{H(f_t),{\bar a}^\dag_t {\bar a}_t\}-H(f_t)\langle {\bar a}^\dag_t {\bar a}_t\rangle_t\right]|\psi_t\rangle \\ \nonumber
&-dt\, g r_1 \langle\psi_t|\left[\frac{1}{2}\{H(f_t),{\bar a}_t {\bar a}^\dag_t\}-H(f_t)\langle {\bar a}_t {\bar a}^\dag_t\rangle_t\right]|\psi_t\rangle \\ \nonumber
&+dN_t^-\left(\frac{\langle {\bar a}_tH(f_t){\bar a}^\dag_t\rangle_t}{\langle {\bar a}_t {\bar a}^\dag_t\rangle_t}-\langle H(f_t)\rangle_t-\omega\right) \\
&+dN_t^+\left(\frac{\langle {\bar a}^\dag_tH(f_t){\bar a}_t\rangle_t}{\langle {\bar a}^\dag_t {\bar a}_t\rangle_t}-\langle H(f_t)\rangle_t+\omega\right).
\end{align}
The work along a quantum trajectory during the time interval $[0,t)$ is the stochastic integral of $dW_t$,
\begin{align}\label{eq:work2}
W_t[\psi]&=\int_0^t dW_s[\psi] \\ \label{eq:work}
&=\Delta U_t[\psi]-Q_t[\psi].
\end{align}
Remarkably, the average of Eq.~\ref{eq:work2} over all realizations, after substituting in Eqs.~\ref{eq:dN} and \ref{eq:dW}, has the simple form
\begin{equation}\label{eq:EWork}
E[W_t]=\int_0^t dt\,{\dot f}_t E[\langle\psi_t|\partial_fH(f_t)|\psi_t\rangle],
\end{equation}
reminiscent of the definition of work for classical systems.
Equation~\ref{eq:EWork} resolves an ambiguity in a common definition of the average work along nonequilibrium processes~\cite{Kieu2004}.
One method for identifying the work is to differentiate the average energy $u_t={\rm Tr}[H(f_t)\rho_t]$ with time,
\begin{equation}\label{eq:diffU}
\partial _tu_t={\rm Tr}[\partial_tH(f_t)\rho_t]+{\rm Tr}[H(f_t)\partial_t\rho_t],
\end{equation}
and then to identify the average work as $w_t={\rm Tr}[\partial_tH(f_t)\rho_t]$ and the average heat as $q_t={\rm Tr}[H(f_t)\partial_t\rho_t]$.
However, when the trace is evaluated using a time-dependent basis, Eq.~\ref{eq:diffU} is no longer correct and the seperation of $\partial_tu_t$ into two parts is not unique~\cite{Esposito2006}.
Equation~\ref{eq:EWork} avoids this ambiguity, as it is evaluated using state vectors instead of density matrices.

\begin{figure}[htb]
\includegraphics[scale=0.65,angle=180]{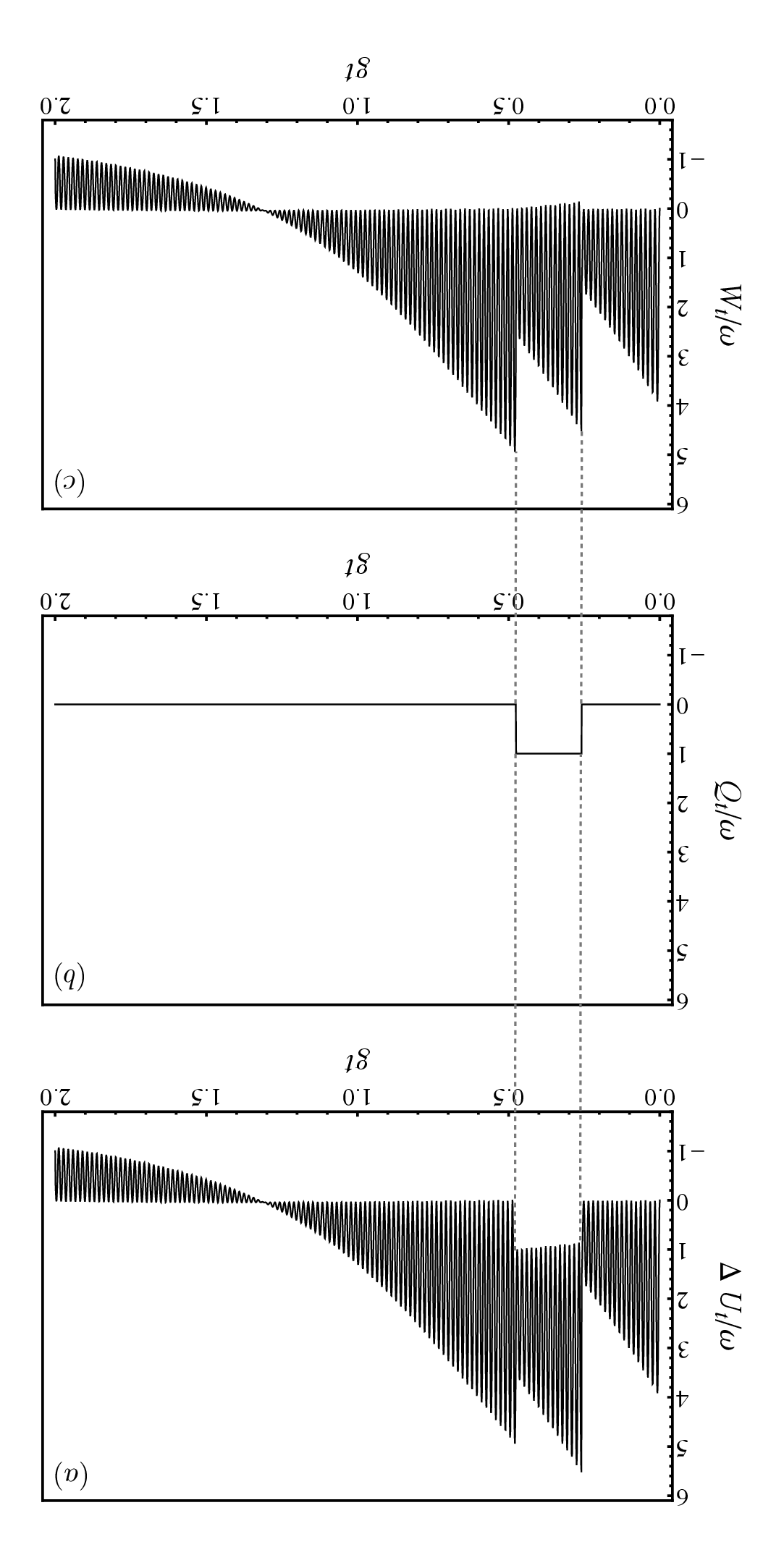}
\caption{Plots of $(a)$ change in energy $\Delta U_t$, $(b)$ heat $Q_t$, and $(c)$ work $W_t$ as a function of time $t$ from $t=0$ to $\tau=80s$ along a representative quantum trajectory with force protocol $f_t=\sqrt{2/(m\omega)}\nu t$ initially in energy eigenstate $|2_{f_0}\rangle$, whose final state is measured to be $|1_{f_\tau}\rangle$.
The additional work required to lower the energy zero, ${\emph w}_t=-\omega\nu^2t^2$, has been neglected.
The parameter values are $\omega=\nu=10\, Hz$, $\lambda = 0.5\, Hz$, $\delta t=0.1\, s$, $g=0.025\, Hz$, $r_1/r_0=0.75$, and $m=1\, kg$.}
\label{fig:traj}
\end{figure}

To illustrate the relationship between work, heat, and energy, I have numerically integrated the stochastic Schr\"odinger equation [Eq.~\ref{eq:sse}] for a thermodynamic process where the initial energy $A=H(f_0)$ and final energy $B=H(f_\tau)$ are measured~\cite{Breuer}.
Plotted in Fig.~\ref{fig:traj} is the change in energy $\Delta U_t$ [Eq.~\ref{eq:energy}], heat $Q_t$ [Eq.~\ref{eq:Q}], and work $W_t$ [Eq.~\ref{eq:work2}] as a function of time from $t=0$ to $\tau=80 s$ along a representative quantum trajectory beginning in energy eigenstate $|2_{f_0}\rangle$ and found at $\tau$ in state $|1_{f_\tau}\rangle$.
The work to lower the energy zero, ${w}_t=-(1/2)m\omega^2\int_0^tds\, \partial_s f_s^2=-\omega\nu^2t^2$, has been neglected in Fig.~\ref{fig:traj} in order to highlight quantum effects.
At two times, $g t\approx 2.6,\, {\rm and}\, 4.9$, there is a discontinuous change in the average energy in Fig.~\ref{fig:traj}$(a)$ when an atom jumps and exchanges a quantum of heat with the reservoir, as depicted in Fig.~\ref{fig:traj}$(b)$.
In between the jumps, the average energy varies rapidly taking on non-integer values, a signature that the oscillator's state is in a coherent superposition of instantaneous energy eigenstates.
This additional variation of the energy is accounted for by the work in Fig.~\ref{fig:traj}$(c)$.
Furthermore, when there are discontinuous jumps the change in the energy is not completely due to the flow  of heat.
As can be seen in Fig.~\ref{fig:traj}$(c)$, the work $W_t$ changes discontinuously as well.
These jumps are a consequence of the oscillator being in a superposition of energy eigenstates.
However, the additional work accrued during these jumps is not work due to the variation of the external parameter $f_t$, as is typical of classical systems.
Its origin is the sudden change in the state vector triggered by the measurement of the outgoing atom, though how this energy is transfered to the oscillator remains unclear.
Nevertheless, this definition of work provides a consistent framework to describe the flow of energy between the oscillator and its environment.

\subsection{Second Law of Stochastic Thermodynamics}\label{sec:entropy}

The second component of a quantum stochastic thermodynamics is a definition of entropy and entropy production along individual quantum trajectories.

I first define the change in entropy of the thermal reservoir.
The role of the thermal reservoir is played by the sequence of atoms; they act as a very large depository for energy with inverse temperature $\beta$.
Specifically, before each thermodynamic process we prepare a large collection of atoms each either in their ground or excited states with relative probability given by the Boltzmann weight~[Eq.~\ref{eq:Boltzmann}].
Over the course of the process, each atom interacts with the harmonic oscillator, but the fraction of atoms that jump is small.
As a result, the relative fraction of atoms in the ground and excited states will not deviate appreciably from the Boltzmann distribution.
Consequently, the change in entropy of the reservoir at time $t$ along a particular quantum trajectory $\psi(\gamma)$ is proportional to the energy absorbed by the thermal reservoir as heat,
\begin{equation}\label{eq:dsres}
ds^r_{t}[\psi]=-\beta dQ_t[\psi].
\end{equation}
Integrating gives the total change in reservoir entropy along the process
\begin{align}\label{eq:delSRes}
\Delta s_{r}[\psi]=\int_0^\tau  ds_{t}^r[\psi]=-\beta Q_\tau[\psi].
\end{align}

In classical stochastic thermodynamics, the entropy of the system is associated with the Shannon entropy of the system's phase space density.
The quantum version of the Shannon entropy is the von Neumann entropy, which at time $t$ reads
\begin{equation}
S(\rho_t)=-{\rm Tr}[\rho_t\ln\rho_t].
\end{equation}
From the definition of the density matrix in Eq.~\ref{eq:rho}, we may rewrite the von Neumann entropy as a classical statistical average over quantum trajectories $\psi(\gamma)$
\begin{equation}\label{eq:vnS}
S(\rho_t)=-E[\langle \psi_t|\ln\rho_t|\psi_t\rangle],
\end{equation}
where here $\rho_t$ is the density matrix for an ensemble of realizations [Eq.~\ref{eq:rho}] and evolves deterministically according to the master equation in Eq.~\ref{eq:master}.
The form of $S$ in Eq.~\ref{eq:vnS} suggests defining the trajectory-depedent system entropy as~\footnote{U.~Seifert, private communication}
\begin{equation}\label{eq:sysEnt}
s_t[\psi]=-\langle\psi_t|\ln\rho_t|\psi_t\rangle.
\end{equation}
Following Seifert~\cite{Seifert2005b}, one may attempt to develop a stochastic differential equation for $s_t$ (as in Eq.~\ref{eq:dU} for $dU_t$); however, there is no obvious compact expression, since $\rho_t$ in general does not commute with its time derivative. 
Over the course of a trajectory that begins in state $|a\rangle$ and ends in state $|b\rangle$ the change in the system's entropy is
\begin{equation}\label{eq:deltaSysEnt}
\begin{split}
\Delta s[\psi]&=-\langle\psi_\tau|\ln\rho_\tau|\psi_\tau\rangle+\langle\psi_0|\ln\rho_0|\psi_0\rangle \\
&=-\ln P_b+\ln P_a,
\end{split}
\end{equation}
where in the second line I used that the density matrix is diagonal at the beginning [Eq.~\ref{eq:rhoA}] and the end [Eq.~\ref{eq:rhoB}] of the process.

Adding Eqs.~\ref{eq:delSRes} and \ref{eq:deltaSysEnt}, we find that the total entropy production during a thermodynamic process is
\begin{equation}
\begin{split}
\Delta s_{tot}[\psi] &=\Delta s[\psi]+\Delta s_{r}[\psi] \\ \label{eq:totEnt}
&= -\ln P_b+\ln P_a-\beta Q_\tau[\psi],
\end{split}
\end{equation}
consistent with the definition proposed by Monnai~\cite{Monnai2005}
In Sec.~\ref{sec:flucRe}, I show that the average total entropy production $\Delta S_{tot}=E[\Delta s_{tot}]$ is a measure of the irreversibility of the thermodynamic process and is non-negative in accordance with the second law of thermodynamics.
Furthermore, the average total entropy production agrees with the definition introduced and analyzed previously by Breuer \cite{Breuer2003}.

\section{Detailed Fluctuation Theorem}\label{sec:flucRe}

In this section, I address the detailed fluctuation theorem~\cite{Bochkov1977a,Bochkov1981a,Gallavotti1995a,Evans2002a,Lebowitz1999,Crooks2000,Hatano2001,Maes2003b,Seifert2005b,Harris2007,Astumian2007} in the context of quantum trajectories.
The detailed fluctuation theorem relates the probabilities to observe particular microscopic trajectories along two thermodynamic processes related by time-reversal.
It has been derived for a wide class of dynamics; in each case,  the detailed fluctuation theorem identifies the source of time-reversal symmetry breaking during a thermodynamic process with the total entropy production~\cite{Maes2000,Maes2003,Maes2003b}.
The generality of this observation suggests that we may use the detailed fluctuation theorem as a tool for ascertaining the entropy production during a nonequilibrium thermodynamic process in situations where its definition may not be obvious.
Through verifying the detailed fluctuation theorem for quantum trajectories, we will see that the source of time-reversal symmetry breaking is the same as the total entropy production $\Delta s_{tot}$ [Eq.~\ref{eq:totEnt}]; thereby, providing evidence that the definition of entropy production in Sec.~\ref{sec:entropy} is consistent with its role as a measure of irreversibility.

The detailed fluctuation theorem for the quantum trajectories of our forced harmonic oscillator relates the probability to observe a quantum trajectory $\psi(\gamma)$ in the forward process $P[\psi]$ to the probability to observe the conjugate trajectory $\tilde\psi(\tilde\gamma)$ in the reverse process ${\tilde P}[\tilde\psi]$ as
\begin{equation}\label{eq:dft}
\ln\frac{P[\psi]}{\tilde{P}[\tilde\psi]}=\Delta s_{tot}[\psi],
\end{equation}
derived below.
In quantum systems, the detailed fluctuation theorem has been demonstrated previously in numerous situations \cite{Monnai2005,Campisi2009,Campisi2010b,Esposito2009,Esposito2007b,Esposito2006,Crooks2008b,DeRoeck2004}.
The essential ingredients are the time-reversal symmetry of the Hamiltonian, and that projective measurements are made at the beginning and at the end of the process.
Moreover, any collection of measurements may be performed during the process without invalidating Eq.~\ref{eq:dft}~\cite{Campisi2010}.
In this respect Eq.~\ref{eq:dft} is not novel; since, the Hamiltonian in Eq.~\ref{eq:H2} is time-reversal invariant; a thermodynamic process, by definition, begins with the measurement of the observable $A$ and is terminated by a measurement of the observable $B$; and the conditional evolution of the state vector $|\psi_t\rangle$ embodied by the stochastic Schr\"odinger equation is merely a sequence of weak measurements.
Nevertheless, I will sketch the derivation of Eq.~\ref{eq:dft}, since alternative approaches sharpen our understanding.

Before deriving Eq.~\ref{eq:dft}, let me comment on one of its consequences.
In particular, notice that Eq.~\ref{eq:dft} immediately implies that $\Delta S_{tot}=E[\Delta s_{tot}]$ equals the relative entropy $D(f||g)=\int dx f(x)\ln[f(x)/g(x)]$ of $P$ and ${\tilde P}$,
\begin{equation}\label{eq:relEnt}
\Delta S_{tot}=D(P||{\tilde P})\ge 0,
\end{equation}
which is always non-negative ($D\ge0$)~\cite{Cover}.
The relative entropy $D(P||{\tilde P})$ is a measure of the distinguishability of $P$ and ${\tilde P}$.
Therefore, $\Delta S_{tot}$ is a measure of  how distinguishable a forward process is from its time reverse~\cite{Callens2004};  it measures the thermodynamic irreversibility of a process~\cite{Maes2003}.
In particular, the equality in Eq.~\ref{eq:relEnt} is reached only for reversible processes where the forward process is indistinguishable from the reverse process, $P[\psi]={\tilde P}[\tilde\psi]$.
Note, unlike the entropy in macroscopic thermodynamics, which enters into the second law of thermodynamics and also encodes information about the properties of equilibrium states such as specific heats, the entropy defined here only reflects the irreversibility of a nonequilibrium thermodynamic process; except  when the thermodynamic process begins and ends in equilibrium.

To verify the detailed fluctuation theorem [Eq.~\ref{eq:dft}], we calculate $P[\psi]$ and ${\tilde P}[\tilde\psi]$.
To determine $P$ and ${\tilde P}$, let me first note that within the quantum trajectory formalism the stochastic evolution of the state vector is the result of a sequence of weak measurements.
Each measurement outcome can be represented by measurement operator~$\Omega$ \cite{Wiseman1996}, whose effect on any pure state $|\psi\rangle$ leads to an unnormalized state vector $|\Psi\rangle=\Omega|\psi\rangle$ that encodes the probability for that outcome in its norm $p(\Omega)=\langle\Psi|\Psi\rangle=\langle\psi|\Omega^\dag\Omega|\psi\rangle$.
Similarly, for a sequence of measurements, the norm of the unnormalized state vector obtained by applying a series of measurement operators equals the probability to observe that sequence of outcomes.
Therefore, the probability to observe any solution of the stochastic Schr\"odinger equation, $P[\psi]$, can be found by determining the norm of the unnormalized state vector resulting from the action of a series of measurement operators describing the effects of observing a series of jumps or no jumps in the sequence of atoms.
These measurement operators can be obtained from the structure of Eq.~\ref{eq:sse}~\cite{Carmichael,Wiseman1994,Jacobs2006,Breuer}.
Below, I simply report the results in order to keep the present discussion concise.

In the forward process, the evolution of the unnormalized state vector in between jumps is realized by an effective time-evolution operator $U_{eff}(t,s)$ --  the solution of
\begin{equation}\label{eq:Ueff}
\partial_tU_{eff}(t,s)=-iH_{eff}(f_t)U_{eff}(t,s),
\end{equation}
with non-hermitian effective Hamiltonian
\begin{equation}\label{eq:Heff}
H_{eff}(f)=H(f)-i\frac{g r_0}{2}{\bar a}^\dag_f{\bar a}_f-i\frac{g r_1}{2}{\bar a}_f{\bar a}_f^\dag
\end{equation}
with initial condition $U(s,s)=I$.
This deterministic evolution is punctuated by discontinuous changes when a measured atom jumps, induced by the jump operators
\begin{equation}
\begin{split}
j_-(t)&=\sqrt{dt\, g r_1 }{\bar a}^\dag_t \\ 
 j_+(t)&=\sqrt{dt\, g r_0 }{\bar a}_t,
\end{split}
\end{equation}
which satisfy
\begin{equation}\label{eq:db}
j_-(t)=j^\dag_+(t)e^{\beta \omega/2}.
\end{equation}
Notably, Eq.~\ref{eq:db}, originally derived by Crooks~\cite{Crooks2008b} in a more general setting assuming the thermal reservoir is in equilibrium, may be viewed as an operator extension of detailed balance to quantum trajectories, relating an atom's transition between energy eigenstates to the energy absorbed by the thermal reservoir as heat.
With this notation, the unnormalized state vector at time $\tau$ conditioned on $\gamma=\{a; m_1,t_1; m_2, t_2; \cdots; m_M,t_M; b\}$ is
\begin{equation}
\begin{split}
|\Psi_\tau(\gamma)\rangle&=|b\rangle\langle b|U_{eff}(\tau,t_N)j_{m_M}(t_M)\cdots j_{m_1}(t_1)U_{eff}(t_1,0)|a\rangle \\ \label{eq:L}
&\equiv |b\rangle\langle b| {\mathcal L}(\gamma)|a\rangle,
\end{split}
\end{equation}
and the probability to observe $\gamma$ given that the initial state is $|a\rangle$ is the norm of $|\Psi_\tau(\gamma)\rangle$,
\begin{equation}\label{eq:Ppsi}
P[\psi|a]=\langle\Psi_\tau(\gamma)|\Psi_\tau(\gamma)\rangle.
\end{equation}

In the reverse process, the force protocol is reversed, implying that $\dot{\tilde f}_t=-\dot{f}_{\tau-t}\propto-\nu$ and that each appearance of $\nu$ in the forward process must be replaced by $-\nu$ in the reverse process.
Consequently, the reverse effective Hamiltonian  at time $t$ that generates the deterministic (but non-hermitian) evolution ${\tilde U}_{eff}(t,s)$, is 
\begin{equation}\label{eq:HeffRev}
{\tilde H}_{eff}({\tilde f}_t)=H({\tilde f}_t)-i\frac{r_0g}{2}{\tilde a}^\dag_t{\tilde a}_t-i\frac{r_1g}{2}{\tilde a}_t{\tilde a}_t^\dag,
\end{equation}
where the reverse displaced raising and lowering operators are obtained from Eq.~\ref{eq:bara} by the substitution $\nu\leftrightarrow -\nu$ as 
\begin{equation}\label{eq:tildea}
\begin{split}
{\tilde a}^\dag_t&=a_{\tau-t}^\dag-i\nu/\omega,\\
{\tilde a}_t&=a_{\tau-t}+i\nu/\omega.
\end{split}
\end{equation}
The discontinuous jumps are induced by reverse jump operators
\begin{equation}
\begin{split}
{\tilde j}_-(t)&=\sqrt{dt\, g r_1 }{\tilde a}^\dag_t \\ 
{\tilde j}_+(t)&=\sqrt{dt\, g r_0 }{\tilde a}_t,
\end{split}
\end{equation}
related to the forward jump operators by
\begin{equation}\label{eq:db2}
j_-(t)=\Theta {\tilde j}_+^\dag(\tau-t)\Theta^{-1} e^{\beta \omega/2}.
\end{equation}

The probability to observe the conjugate  quantum trajectory $\tilde\psi(\tilde\gamma)$ conditioned on initiating the trajectory in state $|{\tilde b}\rangle$ is given by the norm of the unnormalized state vector
\begin{align}
|\tilde\Psi_\tau(\tilde\gamma)\rangle&=|{\tilde a}\rangle\langle {\tilde a}|{\tilde U}_{eff}(\tau,\tau-t_1) \cdots{\tilde j}_{{\tilde m}_1}(\tau -t_N){\tilde U}_{eff}(\tau-t_N,0)|{\tilde b}\rangle \\ \label{eq:LRev}
&\equiv|{\tilde a}\rangle\langle {\tilde a}| \tilde{\mathcal L}(\tilde\gamma)|{\tilde b}\rangle,
\end{align}
as
\begin{equation}\label{eq:PpsiRev}
{\tilde P}[\tilde\psi|{\tilde b}]=\langle\tilde\Psi_\tau(\tilde\gamma)|\tilde\Psi_\tau(\tilde\gamma)\rangle.
\end{equation}

To complete the derivation of Eq.~\ref{eq:dft}, we note that the time reversal invariance of the Hamiltonian, $H(f)=\Theta H(f)\Theta^{-1}$, implies the relationship between effective time-evolution operators,
\begin{equation}\label{eq:timeRev}
U_{eff}(t,s)=\Theta{\tilde U}^\dag_{eff}(\tau-s,\tau-t)\Theta^{-1},
\end{equation}
Combined with Eq.~\ref{eq:db2} this leads to a time-reversal symmetry between ${\mathcal L}$ [Eq.~\ref{eq:L}] and $\tilde{\mathcal L}$ [Eq.~\ref{eq:LRev}]
\begin{equation}
{\mathcal L}(\gamma)=\Theta\tilde{\mathcal L}^\dag(\tilde\gamma)\Theta^{-1} e^{-\beta Q_\tau[\psi(\gamma)]/2},
\end{equation}
which when substituted into the definitions of $P$ [Eq.~\ref{eq:Ppsi}] and ${\tilde P}$~[Eq.~\ref{eq:PpsiRev}], gives
\begin{equation}
\ln\frac{P[\psi|a]}{{\tilde P}[\tilde\psi|{\tilde b}]}=-\beta Q_\tau[\psi].
\end{equation}
Recalling that  in a thermodynamic process the probability for a quantum trajectory to start in state $|a\rangle$ is $P_a$~[Eq.~\ref{eq:rhoA}] and the probability in the reverse process to begin in state $|{\tilde b}\rangle$ is ${\tilde P}_b$ [Eq.~\ref{eq:rhoTilde0}], we find that the log of the ratio of $P[\psi]=P[\psi|a]P_a$ and ${\tilde P}[\tilde\psi]={\tilde P}[\tilde\psi|{\tilde b}]{\tilde P}_b$ is
\begin{equation}\label{eq:dft2}
\ln\frac{P[\psi]}{{\tilde P}[\tilde\psi]}=-\ln {\tilde P}_b+\ln P_a-\beta Q_\tau[\psi]
\end{equation}

Equation~\ref{eq:dft2} is valid for any $P_a$ and ${\tilde P}_b$.
However, as in classical stochastic thermodynamics, there are two noteworthy special cases~\cite{Seifert2005b}.
First, when the initial density matrix of the reverse process is the final density matrix of the forward process, $\tilde\rho_0=\rho_\tau=\rho_B$, we recover Eq.~\ref{eq:dft} equating the total entropy production to the irreversibility of the process.
A second special case of Eq.~\ref{eq:dft2} connects the work dissipated to irreversibility.
Consider a forward process where the initial observable is the initial Hamiltonian, $A=H(f_0)$, the final observable is the final Hamiltonian $B=H(f_\tau)$, and both the forward and reverse processes are started in equilbirium: $\rho_0=\rho^{eq}_{f_0}$ and ${\tilde\rho}_0=\rho^{eq}_{f_\tau}$.
In which case, Eq.~\ref{eq:dft2} reduces to
\begin{equation}\label{eq:dftW}
\ln\frac{P[\psi]}{{\tilde P}[\tilde\psi]}=\beta\left(W_\tau[\psi]-\Delta F\right),
\end{equation}
where $\Delta F$ is the equilibrium free energy difference between the equilibrium ensembles with external parameter values $f_0$ and $f_\tau$.
Thus, the work dissipated -- the work done in excess of the free energy difference -- has a clear physical interpretation as a measure of the irreversibility of a thermodynamic process where the system is driven between two equilibrium states.

To conclude this section, I comment on the work fluctuation relations.
In particular, Eq.~\ref{eq:dftW} immediately leads to a work fluctuation relation relating the probability to observe $W$ work in the forward process $p(W)$ to the probability to observe $-W$ work in the reverse process ${\tilde p}(-W)$~\cite{Crooks1999a}
\begin{equation}
\frac{p(W)}{{\tilde p}(-W)}=e^{\beta(W-\Delta F)},
\end{equation}
and its integral version, the nonequilibrium work fluctuation theorem~\cite{Jarzynski1997a},
\begin{equation}
\left\langle e^{-\beta(W-\Delta F)}\right\rangle=1,
\end{equation}
where here the angled brackets $\langle\cdot\rangle$ denote an ensemble average over work values.
Similar considerations apply to a detailed and integral fluctuation relation for the total entropy production~\cite{Seifert2005b}.

\section{Discussion and Perspectives}\label{sec:discussion}

In this article, I have formulated a quantum stochastic thermodynamics for quantum trajectories.
As a concrete example, I investigated a forced quantum harmonic oscillator.
The oscillator was coupled to a thermal reservoir composed of sequence of two-level atoms, which allowed us to monitor the energy transfer to the atoms and consequently the thermal reservoir leading to a physically motivated definition of heat.
In essence, we are repeatedly measuring the environment and not the system directly.
However, since the environment is broken up into individual quantum systems that interact with the oscillator one at a time, we are able to avoid any practical and conceptual difficulties with measuring an infinite thermal reservoir.
Following the definition of heat, I introduced path-dependent definitions of work and entropy for individual quantum trajectories in Sec.~\ref{sec:stochThermo}.
Their connection with irreversibility was made in Sec.~\ref{sec:flucRe} through the development of a detailed fluctuation theorem for quantum trajectories.

The present discussion focused on a particular model.
Nevertheless, the formulas presented in Sec.~\ref{sec:stochThermo} for work, heat, and entropy could be applied to other systems within the quantum trajectory formalism.
The central issue is whether the measurement scheme can be given a consistent thermodynamic interpretation, as in the present model where the atoms  played the dual role of quantum probe and thermal reservoir.
This opens the interesting question of what other monitoring schemes (so-called unravelings of the master equation), have consistent stochastic-thermodynamic interpretations~\cite{Carmichael,Brun2002,Jacobs2006,Santos2011}.

Within the quantum trajectory formalism, quantum trajectories are the stochastic evolution of the system's state vector through Hilbert space.
A seemingly alternative definition of a trajectory is offered by the consistent (or decoherent) histories~\cite{Griffiths} interpretation of quantum mechanics.
However, Brun has demonstrated that quantum trajectories can be consistently represented within the consistent histories framework~\cite{Brun2000}.
A distinct perspective is obtained by viewing these quantum trajectories as a particular unitary dilation of the damped harmonic oscillator~\cite{Maassen2003}.
A unitary dilation of a quantum Markov semigroup is a representation of the irreversible dynamics as a unitary evolution on a larger Hilbert space, such that when projected onto the system's Hilbert space we recover the original irreversible evolution.
Here, the irreversible dynamics of the harmonic oscillator is dilated onto the unitary dynamics of the oscillator plus the sequence of atoms (see for example Ref.~\cite{Attal2006}).  

A principle motivation for the present study was to develop trajectory-dependent definitions of work, heat, and entropy that could be addressed experimentally.
The quantum harmonic oscillator can be experimentally realized as a single mode of an electromagnetic field in a microwave cavity~\cite{Raimond2001,Walther2006,Gleyzes2007}, and the thermal bath can be engineered by passing a sequence of two-level atoms through the cavity one at a time.
In particular, the work fluctuation relations can be verified experimentally, since it is feasible to construct and measure with a quantum non-demolition measurement the individual energy eigenstates of the electromagnetic field~\cite{Raimond2001}.
The greatest difficulty in verifying the predictions of quantum stochastic thermodynamics is the efficient detection of the atoms once they have interacted with the field.
Quantum stochastic thermodynamics requires a near perfect detection efficiency; whereas, modern experimental setups only reach forty percent~\cite{Raimond2001}.
Nevertheless,  quantum stochastic thermodynamics could be investigated in other quantum systems where the quantum trajectory formalism has been applied, such as quantum dots~\cite{Goan2001}, nanomechanical resonators~\cite{Jacobs2007}, or perhaps in quantum cyclotrons -- where thermal quantum jumps have been observed~\cite{Peil1999}.

Future research directions are manifold.
Particularly interesting is a thermodynamic analysis of effects with a purely quantum origin; such as, the work required to generate entanglement, or the entropy produced in a squeezed thermal bath.
The stochastic Schr\"odinger equation in Eq.~\ref{eq:sse} for the linearly-forced harmonic oscillator may be of interest in its own right.
For example, in the thermodynamic adiabatic limit, Eq.~\ref{eq:sse} could be used to analyze geometric phases in open quantum systems.

\acknowledgements 
I am pleased to acknowledge J.~M.~R.~Parrondo and E.~Lutz for many thought-provoking discussions and suggestions.
Special thanks to T.~Sagawa, M.~Ueda, D.~Porras, A.~Rivas, and U.~Seifert for their comments.
I am particularly grateful to C.~Jarzynski and K.~Jacobs for critically reading this manuscript.
Financial support for this project came from the National Science Foundation (USA) under grant DMR-090660, Grant MOSAICO (Spanish Government), and Grant MODELICO (Comunidad de Madrid).

\appendix

\section{Derivation of the Stochastic Schr\"odinger Equation}\label{sec:derivation}

In this appendix, I derive the stochastic Schr\"odinger equation [Eq.~\ref{eq:sse}] for the time evolution of the one-dimensional forced harmonic oscillator conditioned on continuously monitoring its interaction with a sequence of two-level atoms, by adapting the method used in Refs.~\cite{Ueda1992,Kist1999,Spehner2002}.
The stochastic Schr\"odinger equation is an effective equation of motion that is valid for times long compared to the oscillator-atom interaction time $\delta t$; given that only one atom interacts with the harmonic oscillator at a time; each atom interacts for the same amount of time, $\delta t$; the interaction time is short, $\lambda\delta t\ll1$; the coupling is weak, $\lambda \ll \omega$; and the mean number ${\bar n}_t=\langle {\bar a}^\dag_t {\bar a}_t\rangle_t$ is small $\lambda\delta t{\bar n}_t\ll1$.

The derivation of the stochastic Schr\"odinger equation proceeds by evolving the oscillator-atom state vector, $|\chi_t\rangle$ for a short time from $t_0$ to $t_0+\delta t$ using the Schr\"odinger equation.
We then calculate the probabilities to measure the atom in its ground and excited states in order to demonstrate that  the atom's transitions are described by a Poisson process. 
 
The oscillator-atom state vector at time $t$, $|\chi_t\rangle$, is the solution to the Schr\"odinger equation
\begin{equation}\label{eq:se}
\partial_t|\chi_t\rangle=-i{\mathcal H}(f_t)|\chi_t\rangle,
\end{equation}
where the Hamiltonian 
\begin{equation}
{\mathcal H}(f)=\omega\left(a_f^\dag a_f +\frac{1}{2}\right)-\frac{1}{2}m\omega^2f^2+\omega\sigma^\dag\sigma+\lambda({\bar a}_f^\dag \sigma+{\bar a}_f\sigma^\dag)
\end{equation}
is a sum of $H(f)$ [Eq.~\ref{eq:oscHam}], $H_A$ [Eq.~\ref{eq:atomHam}], and $H_I(f)$ [Eq.~\ref{eq:HI}].
Since each atom is prepared independently from the oscillator, the initial condition at $t_0$ is a factorized state, $|\chi_{t_0}\rangle=|\psi_{t_0}\rangle|\phi_{t_0}\rangle$, with harmonic oscillator state vector $|\psi_{t_0}\rangle$ and atom state vector $|\phi_{t_0}\rangle$.

The analysis is facilitated by first switching to the adiabatic-interaction picture \cite{Klarsfeld1992}.
To this end, let me introduce the translation operator
\begin{equation}\label{eq:A}
A(f)=e^{-ipf},
\end{equation}
whose action on the position operator $x$ and lowering operator ${\bar a}_f$ is given by
\begin{equation}\label{eq:propA}
\begin{split}
&A^\dag(f)xA(f)=x+f,\\ 
&A^\dag(f){\bar a}_fA(f)={\bar a}.
\end{split}
\end{equation}
Using $A$, I introduce the adiabatic-picture state vector
\begin{equation}\label{eq:chiA}
|\chi_t\rangle_A=A^\dag(f_t)|\chi_t\rangle,
\end{equation}
whose equation of motion is found by substituting Eq.~\ref{eq:se} into the time derivative of $|\chi_t\rangle_A$ [Eq.~\ref{eq:chiA}] and exploiting the properties of $A$ in Eq.~\ref{eq:propA},
\begin{equation}\label{eq:adiabatic}
\partial_t|\chi_t\rangle_A=-iH_A(f_t)|\chi_t\rangle_A,
\end{equation}
where the adiabatic Hamiltonian is
\begin{align}
H_A(f_t)=&A^\dag(f_t){\mathcal H}(f_t)A(f_t)+i\dot{A}^\dag(f_t)A(f_t)  \\ 
\label{eq:HA}
=&\omega \left(a^\dag a+\frac{1}{2}\right)-\frac{1}{2}m\omega^2f^2_t +\omega \sigma^\dag \sigma 
\\ \nonumber
&-p{\dot f}_t+\lambda({\bar a}^\dag\sigma+{\bar a}\sigma^\dag).
\end{align}
Next, we shift to the interaction picture using the operator 
$K(t,t_0)$, defined as the solution to the differential equation
\begin{equation}
\begin{split}
\partial_tK(t,t_0)=-i\bigg[\omega \left(a^\dag a+\frac{1}{2}\right)&-\frac{1}{2}m\omega^2f^2_t \\
&+\omega \sigma^\dag \sigma-p{\dot f}_t\bigg]K(t,t_0).
\end{split}
\end{equation}
with initial condition $K(t_0,t_0)=I$.
The solution may be obtained analytically~\cite{Pechukas1966}
\begin{equation}\label{eq:K}
\begin{split}
K(t,t_0)=&\exp\left[-i\omega\left(a^\dag a+\sigma^\dag\sigma+\frac{1}{2}\right)(t-t_0)+\frac{i}{2}m\omega\int_{t_0}^tdx\, f_x^2\right] \\
&\times e^{i\beta_{t,t_0}}D(\alpha_{t,t_0})
\end{split}
\end{equation}
where, recalling that $f_t=\sqrt{2/(m\omega)}\nu t$,
\begin{align}
\alpha_{t,t_0}&=-\sqrt{\frac{m\omega}{2}}\int_{t_0}^t\, dx \dot{f}_xe^{i\omega (x-t_0)} \\
\label{eq:alpha}
&=i\frac{\nu}{\omega}\left(e^{i\omega (t-t_0)}-1\right),\\
\label{eq:beta}
\beta_{t,t_0}&=-\frac{m\omega}{2}\int_{t_0}^t\, dx\int_{t_0}^x\, dy\, \dot{f}_y\dot{f}_x\sin[\omega(y-x)] \\
& = \frac{\nu^2}{\omega}\left\{t-t_0-\frac{1}{\omega}\sin[\omega(t-t_0)]\right\},
\end{align}
and 
\begin{equation}\label{eq:D}
D(\eta)=\exp( \eta a^\dag-\eta^* a)
\end{equation}
 is the displacement operator, whose effect on the lowering operator $a$ is
\begin{equation}\label{eq:displace}
D^\dag(\eta)aD(\eta)=a+\eta.
\end{equation}
The adiabatic-interaction state vector is defined as
\begin{equation}\label{eq:AI}
|\chi_t\rangle_{AI}=K^\dag(t,t_0)|\chi_t\rangle_A.
\end{equation}
The equation of motion for $|\chi_t\rangle_{AI}$ is obtained by differentiating with time Eq.~\ref{eq:AI}, and then substituting in Eqs.~\ref{eq:adiabatic}, \ref{eq:HA}, \ref{eq:K}, \ref{eq:D}, followed by \ref{eq:displace}, to conclude that
\begin{equation}\label{eq:chiAI}
\partial_t|\chi_t\rangle_{AI}=-i H_{AI}(f_t)|\chi_t\rangle_{AI},
\end{equation}
where the adiabatic-interaction Hamiltonian is
\begin{align}
H_{AI}(f_t)&=K^\dag(t,t_0)H_A(f_t)K(t,t_0)+i{\dot K}^\dag(t,t_0)K(t,t_0) \\
\label{eq:HAI}
&=\lambda({\bar a}^\dag\sigma+{\bar a}\sigma^\dag).
\end{align}

The next step is to solve Eq.~\ref{eq:chiAI} perturbatively for short times. 
To clearly identify the approximations involved, I scale time $s=\lambda t$; scale the adiabatic-interaction Hamiltonian ${\tilde H}_{AI}=H_{AI}/\lambda$; and scale the force, $f_t=\sqrt{2/(m\omega)}g_{\nu t}$, by introducing the dimensionless function $g_{\nu t}=\nu t$.
In terms of these scaled quantities, the formal solution of Eq.~\ref{eq:chiAI} --  suppressing the subscripts $AI$ in order to simplify the notation -- 
is
\begin{widetext}
\begin{equation}\label{eq:formSol}
\begin{split}
|\chi_s\rangle=|\chi_{s_0}\rangle
+\sum_{k=1}^\infty(-i)^k\int_{s_0}^s ds_1\,\cdots \int_{s_0}^{s_{k-1}} ds_{k}\, {\tilde H}(g_{\bar\nu s_1})\cdots{\tilde H}(g_{\bar\nu s_k})|\chi_{s_0}\rangle,
\end{split}
\end{equation}
where ${\bar \nu}=\nu/\omega$.
Setting $s=s_0+\delta s$ and expanding Eq.~\ref{eq:formSol} to second order in $\delta s=\lambda\delta t\ll1$, we find
\begin{align}\label{eq:expand1}
|\chi_{s_0+\delta s}\rangle&\sim|\chi_{s_0}\rangle-i\delta s{\tilde H}(g_{\bar\nu s_0})|\chi_{s_0}\rangle-\frac{\delta s^2}{2}\left[{\tilde H}(g_{\bar\nu s_0}){\tilde H}(g_{\bar\nu s_0})+i\partial_s{\tilde H}(g_{\bar\nu s_0})\right]|\chi_{s_0}\rangle, \\
\label{eq:expand2}
&\sim|\chi_{s_0}\rangle-i\delta s({\bar a}^\dag\sigma+{\bar a}\sigma^\dag)|\chi_{s_0}\rangle-\frac{\delta s^2}{2}({\bar a}^\dag\sigma+{\bar a}\sigma^\dag)^2|\chi_{s_0}\rangle
\end{align}
\end{widetext}
where in the second line I substituted in the definition of ${\tilde H}_{AI}$ in Eq.~\ref{eq:HAI}.
The  validity of the asymptotic expansion in Eq.~\ref{eq:expand1} (or Eq.~\ref{eq:expand2}) requires that each successive term is smaller than the previous.
In particular, we must demand that
\begin{equation}\label{eq:asymp1}
\delta s \frac{\nu}{\omega}\ll1.
\end{equation}
Since $\delta s\ll1$, $\nu/\omega$ cannot be too large, restricting the rate at which the force varies.

Now, let us consider the case when the atom is initially in its ground state at $s_0$. 
Substituting $|\phi_{s_0}\rangle=|0\rangle$ into Eq.~\ref{eq:expand2}, leads to
\begin{equation}\label{eq:ground}
\begin{split}
|\chi_{s_0+\delta s}\rangle\sim&|\psi_{s_0}\rangle|0\rangle-i\delta s {\bar a}|\psi_{s_0}\rangle|1\rangle-\frac{\delta s^2}{2}{\bar a}^\dag {\bar a}|\psi_{s_0}\rangle |0\rangle.
\end{split}
\end{equation}
We then measure the state of the atom.
After the measurement, the unnormalized state vector of the harmonic oscillator conditioned on the outcome $|0\rangle$ is obtained by projecting Eq.~\ref{eq:ground} onto $|0\rangle$,
\begin{align}\label{eq:omega0}
|\tilde\psi_0(s_0+\delta s)\rangle \sim|\psi_{s_0}\rangle-\frac{\delta s^2}{2}{\bar a}^\dag {\bar a}|\psi_{s_0}\rangle\equiv \Omega_0|\psi_{s_0}\rangle.
\end{align}
Likewise, when the atom is found to be in the excited state, the unnormalized state vector of the harmonic oscillator conditioned on the outcome $|1\rangle$ is
\begin{align}\label{eq:omega1}
|\tilde\psi_1(s_0+\delta s)\rangle\sim-i\delta s\, {\bar a}|\psi_{s_0}\rangle \equiv \Omega_1|\psi_{s_0}\rangle.
\end{align}

We are interested in time scales long compared to $\delta s$, during which many atoms interact with oscillator.
Therefore, let us consider the time interval $\Delta s=N\delta s$, with $N\gg1$, but the probability for the atom to jump to the excited state remains small $N\delta s^2 {\bar n}_{s_0}\ll1$.
Thus, the probability that more than one jump is observed during $\Delta s$ is negligible.
The probability that no jump occurs during $\Delta s$ is determined from the norm of the unnormalized state vector conditioned on measuring each of a sequence of $N$ atoms in state $|0\rangle$, which according to Eq.~\ref{eq:omega0} is
\begin{align}\label{eq:nojump}
|\tilde\psi_{nj}(\Delta s)\rangle&=\Omega_0^{N}|\psi_{s_0}\rangle \\ \label{eq:nojump2}
&\sim\left(1-\frac{1}{2}N\delta s^2\, {\bar a}^\dag {\bar a}\right)|\psi_{s_0}\rangle.
\end{align}
The probability that no jumps occur is then
\begin{align}
P_{nj}(\Delta s)=\langle\tilde\psi_{nj}(\Delta s)|\tilde\psi_{nj}(\Delta s)\rangle,
\end{align}  
which simplifies to
\begin{align}
P_{nj}(\Delta s)&\sim1-N\delta s^2\langle {\bar a}^\dag {\bar a}\rangle_{s_0}.
\end{align}  
The probability that only one jump is observed at some point during the interval $\Delta s$ is obtained from the unnormalized oscillator state vector conditioned on measuring the $m^{th}$ atom in the excited state
\begin{align}
|\tilde\psi_{j}^m(\Delta s)\rangle&=\Omega_0^{N-m}\Omega_1\Omega_0^{m}|\psi_{s_0}\rangle, \\ \label{eq:jump}
&\sim-i\delta s\, {\bar a}|\psi_{s_0}\rangle,
\end{align}
where in the second line I have substituted in Eq.~\ref{eq:omega1} and retained terms only of order $\delta s$.
Noticeably Eq.~\ref{eq:jump} is independent of $m$.
With the aid of Eq.~\ref{eq:jump}, we find that the probability that one jump occurs at some point during $\Delta s$, 
\begin{align}
P_j(\Delta s)&=\sum_{m=0}^{N-1}\langle\tilde \psi^m_j(\Delta s)|\tilde\psi_{j}^m(\Delta s)\rangle \\ \label{eq:Pj}
&\sim N\delta s^2\langle {\bar a}^\dag {\bar a}\rangle_{s_0}.
\end{align}

We now see from Eq.~\ref{eq:Pj} that the probability to observe a transition during $\Delta s$ is of order $\Delta s=N\delta s$, which is an indication that the series of jumps observed during the evolution of the oscillator is described by a Poisson process.
Moreover, a majority of the time no jump will be observed, and the change in the state vector according to  Eq.~\ref{eq:nojump2} will be small and of order $\Delta s$.
With probability $P_j$ [Eq.~\ref{eq:Pj}] though the atom will jump and the state vector will change dramatically under the action of the lowering operator ${\bar a}$ [Eq.~\ref{eq:jump}].
We may formulate this observation mathematically by introducing a stochastic increment for the Poisson process, $\Delta N_t^+$, which is typically zero in any small time interval $\Delta s=\Delta t/\lambda$, but with probability $P_j$ is one.
Specifically, the Poisson increment is defined by the relations
\begin{equation}
(\Delta N^+_t)^2=\Delta N^+_t
\end{equation}
and 
\begin{equation}
E[\Delta N^+_t]=g\Delta t\langle a^\dag a\rangle_t,
\end{equation}
where I have replaced $s= \lambda t$ and introduced the jump (or decay) rate $g=\lambda^2\delta t$.
Using $\Delta N^+_t$, we may combine Eqs.~\ref{eq:nojump2} and \ref{eq:jump} as
\begin{equation}
\Delta |\tilde\psi_{t_0}\rangle = -\frac{1}{2}g\Delta t\, {\bar a}^\dag {\bar a}|\psi_{t_0}\rangle+\Delta N^+_t ({\bar a}|\psi_{t_0}\rangle-|\psi_{t_0}\rangle),
\end{equation}
which upon taking the infinitesimal limit $\Delta t\to dt$ reads
\begin{equation}\label{eq:dpsi}
d |\tilde\psi_{t_0}\rangle = -\frac{1}{2}gd t\, {\bar a}^\dag {\bar a}|\psi_{t_0}\rangle+d N^+_t ({\bar a}|\psi_{t_0}\rangle-|\psi_{t_0}\rangle).
\end{equation}

To complete the derivation, we recognize that repeating the above sequence of steps with an atom initially in the excited state leads to Eq.~\ref{eq:dpsi} with the replacement ${\bar a}^\dag \leftrightarrow {\bar a}$.
The stochastic Schr\"odigner equation then follows by combining Eq.~\ref{eq:dpsi} and Eq.~\ref{eq:dpsi} with ${\bar a}^\dag \leftrightarrow {\bar a}$, each weighted by the likelihood that an atom is initially in its ground state $r_0$ or an atom is initially in its excited state $r_1$, receptively.
Equation~\ref{eq:sse} is finally recovered by leaving the adiabatic-interaction picture through inverting Eqs.~\ref{eq:adiabatic} and \ref{eq:AI}, and normalizing the state vector over the small interval $dt$.

\section{Rotating Wave Approximation and the Time-Dependent Jaynes-Cummings Interaction Hamiltonian}\label{sec:rotwave}

In this appendix, I argue that the interaction Hamiltonian $H_I(f_t)$ in Eq.~\ref{eq:HI} is a physically relevant interaction.
I will demonstrate that $H_I(f_t)$ is the secular (or rotating wave) approximation of a more general interaction, and well approximates the dynamics on relevant time scales.

An experimental realization of a quantum harmonic oscillator is an electromagnetic field confined to a superconducting microwave cavity~\cite{Raimond2001,Walther2006,Gleyzes2007}.
In such experiments, the force driving the harmonic oscillator corresponds to time-dependent classical macroscopic charges moving determinisiticly.
The two-level systems correspond to atoms traversing the cavity.
Since the atoms are small compared to the wave-length of the electromagnetic field, their interaction is well described in the dipole approximation.
Using the notation for the harmonic oscillator, the atom-field interaction in the dipole approximation in the presence of a classical external field is \cite{Cohen}
 \begin{equation}\label{eq:V}
V(f_t)=\frac{\Omega}{2}(x-f_t)(\sigma^\dag+\sigma),
\end{equation}
where $\Omega$ is the coupling strength, which depends on the dipole moment of the atom.
Substituting in the definitions of $a_t$ and $a_t^\dag$ [Eqs.~\ref{eq:a}], we find
 \begin{equation}\label{eq:V2}
V(f_t)=\lambda(a_t^\dag+a_t)(\sigma^\dag+\sigma),
\end{equation}
where $\lambda=\Omega/\sqrt{2m\omega}$.

By applying a perturbative analysis to the equation of motion for the coupled oscillator and atom, I will verify that when the coupling is weak,
\begin{equation}
\varepsilon=\frac{\lambda}{\omega}\ll1,
\end{equation}
we can ignore the terms in $V$ [Eq.~\ref{eq:V2}] that do not conserve energy ($a_t\sigma$ and $a^\dag_t\sigma^\dag$), and approximate the evolution by $H_I(f_t)$, at the expense of replacing $a_t$ and $a^\dag_t$ by the displaced operators ${\bar a}_t$ and ${\bar a}_t^\dag$

The dynamical evoution of the time-evolution operator $U(t)$ for the coupled oscillator-atom system in the dipole approximation is generated by the sum of $H(f_t)$ [Eq.~\ref{eq:oscHam}], $H_A$ [Eq.~\ref{eq:atomHam}] and $V(f_t)$ [Eq.~\ref{eq:V}] according to the Schr\"odinger equation
\begin{equation}
\partial_tU(t)=-i\left[H(f_t)+H_A+V(f_t)\right]U(t),
\end{equation}
with initial condition $U(0)=I$.
For clarity, we switch to the adiabatic-interaction picture~\cite{Klarsfeld1992}.
To obtain the adiabatic-interaction time-evolution operator, we apply the operators $K^\dag(t,0)$ [Eq.~\ref{eq:K}] and $A^\dag(f_t)$ [Eq.~\ref{eq:A}] to $U$ as
\begin{equation}\label{eq:UAI}
U_{AI}(t)=K^\dag(t,0)A^\dag(f_t)U(t).
\end{equation}
A differential equation for $U_{AI}$ is obtained by differentiating Eq.~\ref{eq:UAI} with time; substituting in the definitions of $K$ [Eq.~\ref{eq:K}] and $A$ [Eq.~\ref{eq:A}]; exploiting their properties in Eqs.~\ref{eq:propA} and \ref{eq:displace}; and finally scaling time $s=\omega t$ to make the $\varepsilon$ dependence explicit:
\begin{equation}\label{eq:UAI2}
\partial_s U_{AI}(s)=-i\varepsilon h(f_s)U_{AI}(s),
\end{equation}
where 
\begin{equation}\label{eq:h}
\begin{split}
h(f_s)=&(a^\dag+\alpha^*_{s,0})\sigma+(a+\alpha_{s,0})\sigma^\dag \\
&+(a+\alpha_{s,0})\sigma e^{-2is}+(a^\dag+\alpha^*_{s,0})\sigma^\dag e^{2is},
\end{split}
\end{equation}
and  $\alpha_{s,0}=i(\nu/\omega)(e^{i\omega s}-1)$ [Eq.~\ref{eq:alpha}].

To solve Eq.~\ref{eq:UAI2} perturbatively using a two-time scale analysis, I introduce a slow time $\tau=\varepsilon t$ and develop an asymptotic expansion for $U_{AI}$ in $s$ and $\tau$,
\begin{equation}\label{eq:pertU}
U_{AI}(s)\sim u^0(s,\tau)+\varepsilon u^1(s,\tau)+\cdots.
\end{equation}
Replacing $\partial_s\to\partial_s+\varepsilon\partial_\tau$ in Eq.~\ref{eq:UAI2} and substituting in Eq.~\ref{eq:pertU}, leads to the differential equation
\begin{equation}\label{eq:pert}
\begin{split}
(\partial_s+\varepsilon\partial_\tau)&[u^0(s,\tau)+\varepsilon u^1(s,\tau)+\cdots]=
\\
&-i\varepsilon h(f_s)[u^0(s,\tau)+\varepsilon u^1(s,\tau)+\cdots],
\end{split}
\end{equation}
which we solve order by order in $\varepsilon$ in order to calculate the terms in the asymptotic expansion of $U_{AI}$ in Eq.~\ref{eq:pertU}.

Equating terms of order $\varepsilon^0$ in Eq.~\ref{eq:pert}, we deduce the differential equation for $u^0$,
\begin{equation}
\partial_s u^0(s,\tau)=0,
\end{equation}
whose solution is
\begin{equation}\label{eq:F}
u^0(s,\tau)=F(\tau),
\end{equation}
where $F(\tau)$ is an unknown function of $\tau$ only, with initial value $F(0)=I$ chosen to satisfy the initial condition $U_{AI}(0)=I$.
We fix $F(\tau)$ by demanding that $\varepsilon u^1$ remain smaller than $u^0$ on times $s=O(\varepsilon^{-1})$ [$\lambda t=O(1)$], maintaining the validity of the asymptotic expansion in Eq.~\ref{eq:pertU} up to times $s=O(\varepsilon^{-1})$.
To this end, we examine the term of order $\varepsilon$ in Eq.~\ref{eq:pert}:
\begin{equation}
\begin{split}
\partial_s u^1(s,\tau)+\partial_\tau u^0(s,\tau)=-i h(f_s)u^0(s,\tau).
\end{split}
\end{equation}
Its formal solution, after substituting in Eqs.~\ref{eq:h}, \ref{eq:F}, and \ref{eq:alpha} is
\begin{equation} \label{eq:u1sol}
\begin{split}
u^1(s,\tau)=&G(\tau)-s\left[i({\bar a}^\dag\sigma+{\bar a}\sigma^\dag)F(\tau)+\partial_\tau F(\tau)\right] \\
&-i \frac{\nu}{\omega}\left[(e^{-is}-1)  \sigma+(e^{is}-1) \sigma^\dag\right]F(\tau) \\
&-i\int_0^s\,dx\left[(a+\alpha_{x,0})\sigma e^{-2ix}+(a^\dag+\alpha_{x,0}^*)\sigma^\dag e^{2ix}\right]F(\tau),
\end{split}
\end{equation}
where $G(\tau)$ is a function of $\tau$ alone with initial condition $G(0)=0$.
Clearly, at times $s=O(\varepsilon^{-1})$, $\varepsilon u^1$ will be the same order of $u^0$, unless we set the term growing lineraly in $s$ to zero:
\begin{equation}
\partial_\tau F(\tau)=-i({\bar a}^\dag\sigma+{\bar a}\sigma^\dag)F(\tau).
\end{equation}
This is a differential equation for $F$ with initial condition $F(0)=I$, whose solution is
\begin{equation}\label{eq:Fsol}
F(\tau)=e^{-i\tau({\bar a}^\dag\sigma+{\bar a}\sigma^\dag)}.
\end{equation}

To lowest order in $\varepsilon$, the approximate solution for $U_{AI}$ is obtained by substituting Eq.~\ref{eq:Fsol} into Eq.~\ref{eq:pertU} and replacing $s=\omega t$, 
\begin{align}\label{eq:Usol}
U_{AI}(t)\sim e^{-i\lambda t({\bar a}^\dag\sigma+{\bar a}\sigma^\dag)},
\end{align}
valid up to times
\begin{equation}
\varepsilon s=\lambda t=O(1).
\end{equation}
An identical expression to Eq.~\ref{eq:Usol} for $U_{AI}$ would be obtained starting from $H_I(f_t)$.
Therefore, the full evolution of $U$ can be approximated up to times $\lambda t=O(1)$ using $H_I(f_t)$ as long as $\lambda/\omega \ll1$.

\bibliography{FluctuationTheory,PhysicsTexts,QuantumWork}

\end{document}